\documentstyle[12pt]{article}

\input psfig
\newcommand{\scite}[1]{$^{\cite{#1}}$}
\def\ltap{\raisebox{-.55ex}{\rlap{$\sim$}} \raisebox{.4ex}{$<$}}
\def\gtap{\raisebox{-.55ex}{\rlap{$\sim$}} \raisebox{.4ex}{$>$}}
\def\gsim{\mathrel{\gtap}}
\def\lsim{\mathrel{\ltap}}
\newcommand{\bm}[1]{\mbox{\boldmath $#1$}}
\def\e{\mbox{e}}
\def\half{{1 \over 2}}
\def\k{\mbox{\bf k}}

\def\p{\mbox{\bf p}}
\def\x{\mbox{\bf x}}
\begin{document}
\begin{flushright}
PURD-TH-97-06 \\
August 1997  \\
hep-ph/9708313
\end{flushright}
\vspace{0.4in}
\begin{center}
{\Large
Non-linear dynamics of reheating}\footnote{
Invited talk at the conference ``Strong and Electroweak Matter '97'',
21--25 May 1997, Eger, Hungary. To appear in the Proceedings.} 
\vspace{0.3in} \\
S. Khlebnikov
\vspace{0.1in} \\
{\it Department of Physics, Purdue University,
West Lafayette, IN 47907, USA} 
\vspace{0.4in} \\
{\bf Abstract}
\end{center}
Resonant decay of the inflaton produces highly
non-equilibrium states dominated by interacting classical
waves. We discuss several topics in the theory of formation
and evolution of such states.
\vspace{0.2in} 
\newpage
\section{Introduction}
Inflationary cosmology\scite{reviews} explains 
the oldness and  the large-scale uniformity
of the observable universe. It may also explain the origin of the
large-scale structure (galaxies, clusters of galaxies). 
It assumes the existence of a separate epoch in 
the cosmological history, when the expansion of the universe
was extremely fast, ``superluminal''. That epoch is called {\em 
inflation}.

Inflation ends through a process known as {\em reheating}, in which  
the energy is transferred from a homogeneous scalar field, the inflaton, 
to other fields and particles. In inflationary cosmology, all the matter 
in our universe has been created as a result of this process.

Reheating can be 
viewed as amplification of quantum fluctuations of various fields
in the time-dependent inflaton background. In slow-roll 
inflation,\scite{reviews} that background is an oscillating homogeneous
inflaton field. It can be viewed alternatively as a Bose condensate of 
zero-momentum particles---the inflaton quanta. Then, reheating is the
decay of the Bose condensate.

It now appears probable that the postinflationary
universe had been a much livelier place than it was 
thought to be. In many models of inflation, the decay of the inflaton  
condensate starts out as a rapid, explosive process, called
preheating,\scite{KLS} during which fluctuations of Bose
fields coupled to the inflaton grow quasiexponentially 
and achieve large occupation numbers. 
This growth can be thought of as a result of parametric 
resonance.\scite{KLS,para,Betal}
At the subsequent stage, called semiclassical thermalization,\scite{us}
the resonance smears out, and the fields reach a slowly evolving 
turbulent state with smooth power spectra.\scite{us,wide,resc,grav,PR}

The highly non-equilibrium states produced in the inflaton decay can
support non-thermal symmetry restoration\scite{effects}
and baryogenesis.\scite{effects,bau} Collisions of the classical waves
produced in the decay process give rise to a strong, conceivably 
detectable background of relic gravitational waves.\scite{grav}

Below, I discuss several topics in the theory of the 
resonant decay of the inflaton and in the description of the resulting
non-equilibrium states. This discussion is based upon partially
published\scite{us,wide,resc,grav} work that I have done in collaboration 
with I. Tkachev. 
 
 \section{A model}
 \label{sect:model}
  Consider the class of models in which the inflaton $\phi$ 
 interacts with another scalar field $X$. The Lagrangian of the
 matter fields is 
 \begin{equation}
L={1\over 2} g^{\mu\nu}\partial_{\mu}\phi \partial_{\nu}\phi
+ {1\over 2}
g^{\mu\nu}\partial_{\mu} X \partial_{\nu} X - V(\phi,X)
\label{L}
\end{equation}
 where 
\begin{equation}
V(\phi,X)=V_{\phi}(\phi)  + {1\over 2} g^2 \phi^2 X^2
+{1\over 2} M_{X}^{2} X^{2}
\label{pot}
\end{equation}
and $V_{\phi}$ is a potential for the field $\phi$. 
The space-time metric $g_{\mu\nu}$ is assumed to be that 
of a spatially flat Friedmann-Robertson-Walker (FRW) space-time
with the scale factor $a(t)$. In our simulations, the scale
factor can be determined self-consistently, using the Einstein
equations. 

For definiteness, let us now consider the so-called massive case, when
\begin{equation}
V_{\phi}(\phi)={1\over 2} m^{2} \phi^{2}
\label{m2}
\end{equation}
Later, we will consider other potentials for $\phi$.
If we want our simple model to be a realistic model of inflation,
the inflaton mass $m$ should be of order $10^{-6} M_{\rm Pl}$,
to comply with the COBE data.

Both $\phi$ and $X$ are regarded (for a while) as quantum operators, 
and we are working in the Heisenberg picture: the time 
dependence of averages is due to the time-dependence of operators, while 
the quantum state is time-independent. The system is translationally 
invariant in space. 
So, the average of the inflaton field $\phi$ in the quantum state is
space-independent:
 \begin{equation}
\langle \phi(\x,t) \rangle=\phi_{0}(t)
\label{ave}
\end{equation}
We consider the initial data problem formulated as follows.

The inflaton's zero-momentum (homogeneous) mode $\phi_{0}$
rolls slowly during inflation, until it reaches a certain value, 
of order $M_{\rm Pl}/3$. At that time, the slow roll changes into
oscillations of $\phi_{0}$ about the minimum of $V_{\phi}$. Thus
the inflation ends and the reheating starts.\scite{reviews}
The inflaton's homogeneous mode decays into fluctuations of $X$
and those of its own field $\phi$.
The change from the slow-roll to the oscillatory behavior of $\phi_{0}$
is not instantaneous, and in principle the method described below
allows us to follow the evolution of the fluctuations through the 
inflationary stage and into the reheating. The results, however, are
practically unchanged if we adopt, instead, the following simpler
formulation.

To begin with, we solve for $\phi_{0}$ and the scale factor $a$
at the last stages of inflation and during the crossover from the 
slow roll to the oscillating stage. For this purpose, the fluctuations 
can be neglected. The crossover solution is insensitive to 
the values of $\phi_0$ and $d\phi_0/dt$ with which the slow roll starts.
It turns out that the solution for $\phi_{0}$, as a function of 
the {\em conformal} time $\eta$, $d\eta=dt/a(t)$, has an extremum
during the crossover, before it makes the first oscillation. We will 
refer to the extremum
as the time when the inflaton oscillations start; we take that time
as $\eta=0$. In the massive model, 
$\phi_{0}(0)=0.28 M_{\rm Pl}$.\scite{wide} 
The fluctuations at $\eta=0$ are small, and in our formulation we 
assume that they are
in the conformal vacuum, with respect to annihilation operators 
defined below. This specifies the (quantum) initial
conditions for the fluctuations. 
(The assumption of the conformal vacuum is not a crucial one, as 
long as it gives initial fluctuations of the correct order of 
magnitude.\scite{us})

 It is convenient to rescale to the dimensionless conformal 
 variables: 
 \begin{center}
\begin{tabular}{ll}
$\tau  =  m \eta\; ;$ &  ~~~~~~~$\varphi= \phi a/\phi_{0}(0)$ \\
$\mbox{\boldmath $\xi$} =m \mbox{\bf x}\; ;$  
& ~~~~~~~$\chi=X a/\phi_{0}(0)$
\end{tabular} 
\end{center}
We have normalized the scale factor $a(\tau)$ to 1 at $\tau=0$, when the
inflaton oscillations start: $a(0)=1$.
By virtue of our rescaling, $\varphi_{0}(0)=1$.
The action for the matter fields in the new variables takes the form 
\begin{eqnarray}
S & = & \frac{\phi_{0}^{2}(0)}{m^{2}}  \int d^{3}\xi d\tau 
      \left[ \half ( \dot{\varphi}- h \varphi )^{2}
           +\half ( \dot{\chi}- h \chi )^{2}
           - \frac{(\nabla_{\xi} \varphi)^{2}}{2} 
           - \frac{(\nabla_{\xi} \chi)^{2}}{2} 
               \right. \nonumber\\  
  & &    \left.  -\half a^{2} \varphi^{2}
           - 2q \varphi^{2} \chi^{2}
           -\half \frac{M_{X}^{2}}{m^{2}} a^{2} \chi^{2}
        \right]
\label{Sconf}
\end{eqnarray}
Hereinafter dots denote derivatives with respect to $\tau$;
$h=\dot{a}/a$ is the rescaled Hubble parameter. 

The parameter $q$ appearing in (\ref{Sconf}) is defined as
\begin{equation}
q=\frac{g^{2}\phi^{2}_{0}(0)}{4m^{2}}
\label{q}
\end{equation}
and is called the resonance parameter. 
 
In this type of inflationary models, the resonance parameter $q$ is 
naturally large.\scite{KLS} Indeed, with 
$\phi_{0}^{2}(0)\sim 0.1 M_{\rm Pl}^{2}$ and 
$m^{2}\sim 10^{-12} M_{\rm Pl}^{2}$, one gets $q\sim 10^{10}g^{2}$,
so even for fairly small $g^{2}$ one still has $q\gg 1$. A large value
of $q$ is necessary for efficient resonant decay of the inflaton
in this model.\scite{KLS,wide}

The Heisenberg equations of motion following from (\ref{Sconf})
are
\begin{eqnarray}
\ddot{\varphi}-\nabla_{\xi}^{2} \varphi -\frac{\ddot{a}}{a}\varphi
+a^{2} \varphi + 4q \varphi \chi^{2} & = & 0 
\label{eqphi} \\
\ddot{\chi}-\nabla_{\xi}^{2} \chi  -\frac{\ddot{a}}{a}\chi +
m_{\chi}^2 a^{2} \chi + 4q \varphi^{2} \chi & = & 0 
\label{eqchi}
\end{eqnarray}
where $m_{\chi}\equiv M_X/m$. We can write
\begin{equation}
\varphi(\bm{\xi},\tau)=\varphi_{0}(\tau) +\delta\varphi(\bm{\xi},\tau)  
\label{delta}
\end{equation}
where $\delta\varphi(\bm{\xi},\tau)$ is the (quantum) fluctuation. Recall 
that $\phi_{0}(\tau)$ defined in (\ref{ave}) and, consequently,
$\varphi_{0}(\tau)$ are $c$-number functions.

\section{Quantities to compute}
What are the interesting quantities to compute? One is the power
spectrum of  fluctuations. This is defined as follows.
Introduce the spatial Fourier components
\begin{equation}
\chi(\bm{\xi},\tau)=\int \frac{d^{3} k }{(2\pi)^{3/2}} ~\chi_{\k}(\tau) 
\e^{i\k\bm{\xi}} \; ,
{}~~~~~\delta\varphi(\bm{\xi},\tau)=
\int \frac{d^{3} k }{(2\pi)^{3/2}} ~\varphi_{\k}(\tau) \e^{i\k\bm{\xi}}
\label{four}
\end{equation}
Thus $\k$ is a {\em comoving momentum} (in the units of the mass $m$); 
it does not redshift in an expanding universe. The power spectra
are then defined as
\begin{equation}
P_{\chi}(k) = \langle|\chi_{\k}|^{2} \rangle/V \; ,
{}~~~~~P_{\varphi}(k) =
 \langle |\varphi_{\k}|^{2} \rangle/V
 \label{pwsdef}
 \end{equation}
 where $V$ is a normalization spatial volume. Because the system 
is isotropic, the power spectra are functions only of $k=|\k|$. 
 
 The angular brackets so far denote 
 averaging over the quantum state of the system.
 The actual calculations were done using the classical approximation;
 we will come to this important point shortly. In the classical 
 approximation, the averages over the quantum state are replaced by
 averages over realizations of the random initial data.\scite{us}
 
 Another interesting quantity is the variance of a field. With the
 above definitions, it would be just an integral over $d^{3}k$ of the 
 field's power spectrum, if we did not have to subtract away the 
 high-momentum quantum fluctuations in order
 to make the variance finite.
 A natural scheme is to subtract the full initial value of the variance:
 \begin{equation}
 \langle \chi^{2}\rangle_{1}(\tau)=
 \int [P_{\chi}(k,\tau)-P_{\chi}(k,0)] d^{3}k
 \label{sub1}
 \end{equation}
 If we introduce a maximal (cutoff) momentum $k_{\max}$, we can
 view (\ref{sub1}) as the limit of 
\begin{equation}
\langle \chi^{2} \rangle_{2}(\tau)=\int_{k<k_{\rm max}}
P_{\chi}(k,\tau) d^{3} k -
\int_{k<k_{\rm max}} P_{\chi}(k,0) d^{3}k
\label{sub2}
\end{equation}
as  $k_{\rm max}$ goes to infinity.
A maximal momentum is automatically present in our lattice 
calculations. We have to make sure that it is large enough
so that (\ref{sub2}) is insensitive to the cutoff or, equivalently,
the modes with $k\sim k_{\rm max}$ do not give a large contribution
to (\ref{sub2}).

The variances measure the typical size of the fluctuations. For
example, they determine the fluctuations' symmetry restoring 
power.\scite{effects} Of main interest are the values of the variances
at sufficiently large times, when the fluctuations are the largest.
For the values of $k_{\rm max}$ that we use,
the first term on the right-hand side of (\ref{sub2}), when resonance
is efficient, quickly becomes much larger than the second.
So, we will ``forget'' the second term and use
\begin{equation}
\langle \chi^{2} \rangle(\tau)=\int_{k<k_{\rm max}}
P_{\chi}(k,\tau) d^{3} k
\label{var}
\end{equation}
The parameters of our model ($m$, $M_{X}$, $g^{2}$) can 
be viewed as the running parameters of quantum field theory,
normalized at the momentum point $k_{\rm max}$.

In our actual calculations, the 
variances are obtained as lattice averages:
\begin{eqnarray}
\langle \chi^{2} \rangle & = & 
{1\over N_s} \sum_{i=1}^{N_s} \chi_{i}^{2}
\label{varlat} \\
\langle (\delta\varphi)^{2} \rangle & = & 
{1\over N_s} \sum_{i=1}^{N_s}
\varphi_{i}^{2}  - 
\left( {1\over N_s} \sum_{i=1}^{N_s} \varphi_{i} \right)^{2}
\label{latphi}
\end{eqnarray}
where $i$ labels the lattice sites, and $N_s$ is their total number; 
\begin{equation}
{1\over N_s} \sum_{i=1}^{N_s} \varphi_{i} =\varphi_{0}
\label{zeromode}
\end{equation}
is the inflaton's zero-momentum mode.

Note that in the classical approximation, 
the variances (\ref{varlat}), (\ref{latphi}) and the zero mode 
(\ref{zeromode}) need not be averaged over realizations of 
the initial data. These quantities are
self-averaging, as they are defined as averages over a large number 
lattice sites. Similarly, for the power spectra in the classical 
approximation, we use, instead of (\ref{pwsdef}),
\begin{equation}
P_{\chi}(k) = \frac{1}{V}\int |\chi_{\k}|^{2} \frac{d\Omega}{4\pi} \; ,
{}~~~~~P_{\varphi}(k) =
 \frac{1}{V}\int |\varphi_{\k}|^{2} \frac{d\Omega}{4\pi}
 \label{pws}
 \end{equation}
where $\Omega$ is the direction of $\k$.

\section{The linear stage}
As seen from (\ref{Sconf}), the parameter 
$m^{2}/\phi_0^{2}(0)\sim 10^{-11}$ suppresses the initial magnitude 
of the fluctuations relative to the scale at which their 
non-linear interactions become important. Unless the parameter
$q$ is {\em too} large, in which case the scale of non-linearity is itself 
too small, the initial evolution of the fluctuations proceeds in the linear
regime.

Expanding the equations (\ref{eqphi})--(\ref{eqchi}) to the first
order in $\delta\varphi$ and $\chi$, we obtain the following
{\em linearized equations}
\begin{eqnarray}
\ddot{\varphi}_{0}-(\ddot{a}/a)\varphi_{0}
+ a^{2} \varphi_{0} & = & 0 
\label{lphi} \\
\delta\ddot{\varphi}-(\ddot{a}/a)\delta\varphi+ 
a^{2} \delta\varphi & = & 0 
\label{ldphi} \\
\ddot{\chi}-\nabla_{\xi}^{2} \chi  -(\ddot{a}/a)\chi +
m_{\chi}^{2} a^{2} \chi 
+ 4q \varphi_{0}^{2} \chi & = & 0 
\label{lchi}
\end{eqnarray}
$m_{\chi}\equiv M_{X}/m$. 
The Fourier expansion (\ref{four}) turns the equation (\ref{lchi})
into
\begin{equation}
\ddot{\chi}_{\k} + \omega^{2}_{k}(\tau) \chi_{\k} = 0
\label{equ}
\end{equation}
where
\begin{equation}
\omega^2_k(\tau) = m_\chi^2 a^2(\tau)+k^2 - {\ddot a}/{a}
+4q\varphi_0^2(\tau)  
\label{ome}
\end{equation}
Eq. (\ref{equ}) is still a quantum 
equation, but it is a linear equation and can be solved exactly.

The operator solution to (\ref{equ}) can be written as
\begin{equation}
\chi_{\k}(\tau) = 
f_{k}(\tau) b_{\k}(0) + f^*_{k}(\tau) b^{\dagger}_{-\k}(0)
\label{ope}
\end{equation}
The $c$-number function $f_{k}(\tau)$ satisfies the same
equation as $\chi_{\k}$,
\begin{equation}
\ddot{f}_{k} + \omega^{2}_{k} f_{k}= 0
\label{eqf}
\end{equation}
 and the initial conditions
\begin{eqnarray}
f_{k}(0) & = & \frac{m}{\phi(0) \sqrt{2\tilde{\omega}_{k}(0)}}
\label{ini} \\
{\dot f}_{k}(0) & = &
\left[ -i\tilde{\omega}_k(0) + h(0) \right] f_k(0) 
\label{dotini}
\end{eqnarray}
where
\begin{equation}
\tilde{\omega}^{2}_{k}(\tau)=m_\chi^2 a^2(\tau)+k^2 
+4q\varphi_0^2(\tau)  =\omega^2_k(\tau) +{\ddot a}/{a}
\label{tilome}
\end{equation}
Note that $f_k$ depends only on the absolute value of $\k$.
The operators $b$ and $b^{\dagger}$ are time-independent creation 
and annihilation operators normalized according to
\begin{equation} 
[b_{\k},b^{\dagger}_{\k'}]=\delta(\k-\k')
\label{com}
\end{equation}
The quantum state of the system is assumed to be the vacuum 
of $b_{\k}$. The time-dependence of the operators $\chi_{\k}$ is entirely 
due to that of the $c$-number mode functions $f_{k}$.

\section{Non-adiabatic amplification}
In the linear approximation, the evolution of the inflaton 
zero-momentum mode is given by (\ref{lphi}). In a static universe,
i.e. for $a(\tau)=1$, the relevant solution to (\ref{lphi}) is 
\begin{equation}
\varphi_{0}(\tau)=\cos\tau
\label{sin}
\end{equation} 
Then, the equation (\ref{eqf})  reduces to the well-known 
Mathieu equation:
\begin{equation}
\ddot{f_{k}}+ [A+2q \cos 2\tau] f_{k} = 0 
\label{mathieu}
\end{equation}
where $A=k^{2}+m_{\chi}^{2}+2q$. The instability zones of the
Mathieu equation, in the plane of $A$ and $q$,
determine momenta $k$ for which there are exponentially growing
solutions. This growth is known as {\em parametric resonance}.
The case of large $q$ is known in the literature as the broad
resonance case.\scite{KLS}

\begin{figure}
\psfig{file=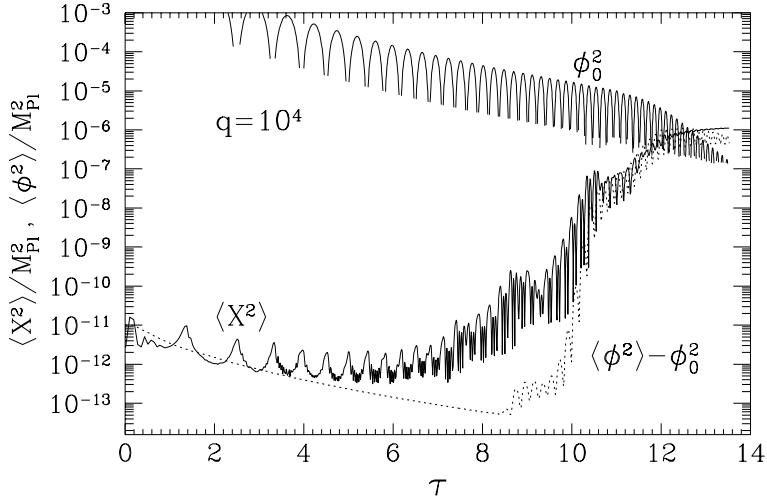,height=7cm,width=10.5cm}
\caption{Variances of the physical fields $X$ and $\phi$, together
with the inflaton's zero-momentum mode, in the massive
model with $q=10^4$, $M_X=0$ in the expanding universe.}
\label{xi2_phi2_qe4_exp}
\end{figure}

In the expanding universe, solutions to (\ref{lphi}) are not periodic.
The crucial point is that, when $q$ is sufficiently large, there still are 
growing solutions (although the growth is no longer simply exponential),
see Fig. \ref{xi2_phi2_qe4_exp}.
 The range of $q$ for which there is an efficient growth of the variance
 of $X$ has been determined numerically.\scite{wide} For 
example, for $M_{X}=0$, it is $q\gsim 10^{3}$.
The growing solutions in the expanding universe have been recently 
explored analytically.\scite{towards}

To distinguish the case with a non-periodic
$\varphi_{0}$ from the case of a periodic $\varphi_{0}$, we will use,
instead of the term parametric resonance, the term {\em
non-adiabatic amplification}. This stresses that the periodicity of
$\varphi_{0}$ is not crucial for the existence of the growing solutions.
What is crucial for the efficient amplification is that, at least once in a while, 
the adiabatic  (WKB) condition for Eq. (\ref{equ})
\begin{equation}
\frac{|\dot{\omega}_{k}(\tau)|}{\omega_{k}^{2}(\tau)} \ll 1
\label{adi}
\end{equation}
breaks down.

Soon after the inflaton oscillations start, the inflaton zero-momentum 
mode is well approximated by
\begin{equation}
\varphi_{0}(\tau) \approx {\bar \varphi}(\tau) \cos(a\tau + \alpha)
\label{barvar}
\end{equation}
where ${\bar \varphi}$ is a relatively slowly
decreasing amplitude. During the linear stage, its decrease
is entirely due to the redshift of the field:
\begin{equation}
 {\bar \varphi}(\tau) \sim a^{-1/2}(\tau)
 \label{dec}
 \end{equation}
The amplitude of the physical field $\phi_{0}$ is 
${\bar \phi}(\tau)=\phi_{0}(0){\bar \varphi}(\tau)/a(\tau)$.

It is convenient to introduce the effective resonance 
parameter\scite{wide}
\begin{equation}
q_{\rm eff}(\tau)=\frac{g^{2}{\bar \phi}^{2}(\tau)}{4m^{2}}
=q\frac{{\bar \varphi}^{2}(\tau)}{a^{2}(\tau)}
\label{red}
\end{equation}
For example, in Fig. \ref{xi2_phi2_qe4_exp}, at $\tau\sim 10$,
we have $\bar\phi^{2}\sim 10^{-5} M_{\rm Pl}^{2}$, so
$q_{\rm eff}\sim 1$.

The most interesting case is that when the resonance
parameter $q$ is large enough for the amplification to fully develop.
That means that the growth of the amplified modes is cut-off  by 
non-linear effects, rather than by the expansion of the universe.
This is the case when the fluctuations can reach the largest values.
It occurs when at the end of the linear stage the effective
resonance parameter is still fairly large: 
$q_{\rm eff}\gsim \max\{1, (M_{X}/m)^{4}\}$, which translates into
$q\gsim 10^{4} \max\{1, (M_{X}/m)^{4}\}$.\scite{wide}

As long as $q_{\rm eff}\gg 1$, the adiabatic approximation is good 
most of the time. The condition (\ref{adi}) breaks down
only during small intervals when $\varphi_{0}$ is close to zero:
\begin{equation}
\cos(a\tau+\alpha) \lsim  q_{\rm eff}^{-1/4}
\label{cos}
\end{equation}
and, even then, only provided that 
\begin{equation}
k^{2}/a^{2}+m_{\chi}^{2} \lsim q_{\rm eff}^{1/2}
\label{band}
\end{equation}
These intervals of time correspond to spikes in 
$\langle X^{2}\rangle$, which are
seen in Fig. \ref{xi2_phi2_qe4_exp}. We will distinguish between 
$\langle \chi^2\rangle$ in spikes and in ``valleys'' between them
by a subscript: ``s'' or ``v''. 
One can show analytically that during the linear stage
$\langle \chi^2\rangle_{\rm s}/\langle \chi^2\rangle_{\rm v}
\sim q_{\rm eff}^{1/4}$.

The band of momenta specified by (\ref{band}) is analogous to 
the strongest (for a given $q$) instability band of the Mathieu 
equation (\ref{mathieu}). Notice that
$k/a$ is the physical momentum in the units of $m$.
For a given $k$, even one that satisfies (\ref{band}),
the mode function $f_{k}$ will not grow each time the adiabatic 
condition is broken,\scite{wide,towards} so the power spectrum of 
$\chi$ at the linear stage becomes jagged.\scite{resc}

\section{Quantum-to-classical transition}
The modes of $\chi$ with growing $f_{k}$ achieve large occupation 
numbers and begin to behave classically.
The theory of this  phenomenon is parallel to the linear theory\scite{PS} 
describing how large-scale classical density perturbations are 
generated from the vacuum fluctuations {\em during} inflation. 

The occupation number in the modes with $|\k|=k$,
obtained by a Bogoliubov transformation, is
\begin{equation}
n_{k}(\tau) = \frac{\phi_{0}^{2}(0)}{2m^{2}\tilde{\omega}_{k}(\tau)}
\left[ |\dot{f}_{k}-h f_{k}|^{2}(\tau) +
\tilde{\omega}_{k}^{2}(\tau)|f_{k}|^{2}(\tau)  \right]-1/2
\label{n}
\end{equation}
One can show that up to $O(1/n_{k})$ the quantum averages
of the operators in the modes with $|\k|=k$ can be
approximated by classical averages computed with the help
of a certain distribution function. When the state vector
is the vacuum of $b_{\k}$ and $b_{-\k}$ ,
the distribution function for the pair of modes with momenta
$\k$ and $-\k$ is\scite{PS,us}
\begin{equation}
{\cal F}[\chi_{\k}, {\dot \chi}_{\k};\tau] = {\cal N}
  \exp\left( -\frac{(2\pi)^{3} |\chi_{\k}|^2}{V|f_k(\tau)|^2} \right)
\delta
  \left(f_k(\tau) {\dot \chi}_{\k}- {\dot f}_k(\tau) \chi_{\k} \right)
\label{dis}
\end{equation}
where the delta function is a shorthand for the product of the
delta functions for real and imaginary parts;
${\cal N}$ is a time-independent normalization. (Note that half
of the values of $\k$ give all the independent $\chi_{\k}$.)

It is useful to notice that (\ref{dis})
satisfies the classical Liouville equation during the entire linear stage.
So, we can evolve it back and use it as an initial condition at $\tau=0$,
where $f_k$ and its derivative are given by (\ref{ini}),
(\ref{dotini}). This leads to the classical problem with
random initial conditions distributed according to\scite{us,wide}
\begin{equation}
{\cal F}_k[\chi_{\k}, {\dot \chi}_{\k};0] = {\cal N}'
\exp\left(-\frac{2\phi^2(0)}{V' m^2} \tilde{\omega}_k(0)
            |\chi_{\k}|^2 \right)
  \delta
\left({\dot \chi}_{\k}+ [i\tilde{\omega}_{k}(0)-h(0)]  \chi_{\k} \right)
\label{disini}
\end{equation}
$V'\equiv V/(2\pi)^{3}$.
This classical problem and the original quantum problem
``converge'' as the modes are amplified.

How good is this classical approximation?
For example, if $q_{\rm eff}\sim 1$ at the end of the linear stage,
the typical occupation number in the efficiently amplified modes
at that time is $n_{\rm amp}\sim\bar{\phi}^{2}/m^{2}$. The
corrections to the classical approximation are then of order
$1/n_{\rm amp}\sim 10^{-7}$.

When
\begin{equation}
|f_{k}| |\dot{f}_{k}| \sim m^{2} n_{k}/\phi_{0}^{2}(0)
\label{cond}
\end{equation}    
and $n_k\gg 1$, one can 
derive the classical approximation in the following simple way.
The solutions $f_{k}(\tau)$ and $f_{k}^{*}(\tau)$ to 
(\ref{eqf})--(\ref{dotini}) satisfy
\begin{equation}
f_{k}(\tau)\dot{f}_{k}^{*}(\tau) -
f_{k}^{*}(\tau)\dot{f}_{k}(\tau)= (m^{2}/\phi_{0}^{2}(0)) i 
\label{wro}
\end{equation}
Given (\ref{cond}) and $n_k\gg 1$, the right-hand side of (\ref{wro})
can be neglected compared to each of the terms on the left-hand side.
That means that the phase $\theta_k$ defined by
\begin{equation}
f_{k}^{*}=f_{k}\exp(i\theta_{k})
\label{phase}
\end{equation}
becomes almost time-independent. 
With the definition (\ref{phase}), the operator solution (\ref{ope})
takes the form
\begin{equation}
\chi_{\k}(\tau) = f_{k}(\tau) \left[ b_{\k}+\e^{i\theta_{k}} 
b_{-\k}^{\dagger} \right]\equiv f_{k}(\tau) {\cal O}_{\k}
\label{solq}
\end{equation}
When the phase $\theta_k$ is almost time-independent, the time-derivative
of $\chi_{\k}$ can be found as
\begin{equation}
\dot{\chi}_{\k'}\approx \dot{f}_{k'} {\cal O}_{\k'}
\label{dotchi}
\end{equation}
provided $|\dot{f}_{k}|$ is large enough, cf. (\ref{cond}).
The canonical momentum conjugate to $\chi_{\k}$, as obtained from
the action (\ref{Sconf}),  is proportional
to $({\dot \chi}_{-\k}-h\chi_{-\k})$.
A direct calculation shows that $[{\cal O}(\k), {\cal O}(\k')]=0$, 
for any $\k$ and $\k'$.
So, in this approximation, $\chi_{\k}$ and its canonical momentum 
commute and can therefore be regarded as random classical variables.

Even if at some time the condition (\ref{cond}) breaks down, i.e.
one of the two quantities, $|f_k|$ or $|\dot{f}_k|$, happens to be small,
the classical approximation will still work, as long as $n_k\gg 1$.
The reason is that the other of the two quantities is necessarily large, 
cf. (\ref{n}), so the quantum ``noise" will not be able to disturb the evolution 
of the modes.

In the vacuum of $b_{\k}$, the variance 
(\ref{sub1})  is
\begin{equation}
\langle \chi^{2}\rangle_{1}(\tau)=
\int \frac{d^{3} k}{(2\pi)^{3}} |f_{k}|^{2}(\tau)
\label{beard}
\end{equation}
As long as $q_{\rm eff}\gg 1$, $f_{k}$ oscillates rapidly in the
``valleys''. When $\theta_{k}$ becomes almost time-independent,
$|f_{k}|^{2}$ also oscillates rapidly there. As a result, the variance 
acquires a high-frequency ``beard'', see Fig. \ref{xi2_phi2_qe4_exp}.

\section{Rescattering}
The modes outside the band (\ref{band}) and fluctuations of the field 
$\varphi$ do not undergo an efficient non-adiabatic amplification 
in the massive model. The system thus arrives into the regime when 
the low-momentum
fluctuations of $\chi$ are already large, while its higher-momentum
fluctuations and the fluctuations of $\varphi$ are still small.

Fluctuations of $\varphi$ now get amplified
via {\em rescattering}\scite{us} of the already large fluctuations 
in the low-momentum modes of $\chi$. Further rescattering of the
$\chi$ and $\varphi$ fluctuations leads to amplification of the
higher-momentum modes of both $\varphi$ and $\chi$.\scite{resc}

The early stages of these processes can be studied using the following 
approximation. We assume that the fluctuations of $\varphi$ are 
much smaller than the amplitude of its zero-momentum mode:
\begin{equation}
 \langle (\delta\varphi)^{2}\rangle\ll \bar{\varphi}^{2}
\label{ll}
\end{equation}
but we make no assumptions about the size of the fluctuations 
of $\chi$. Expanding the full equations (\ref{eqphi}), (\ref{eqchi})
in $\delta\varphi$ (but no longer in $\chi$),
we obtain, to the first order, the following approximate equations 
($\p\neq 0$)
\begin{eqnarray}
{\ddot \chi}_{\k}+\omega^{2}_{k}(\tau)\chi_{\k}+
8q\varphi_{0}(\tau) \int d^{3}p \varphi_{\p}^{*} \chi_{\k+\p} 
& = & 0 \label{appchi} \\
{\ddot \varphi}_{\p} + \Omega_{p}^{2}(\tau) \varphi_{\p}+
4q \varphi_{0}(\tau) \int d^{3}k \chi_{\k}^{*} \chi_{\k+\p} 
& = & 0 \label{appphi} \\
{\ddot \varphi}_{0} - ({\ddot a}/a) \varphi_{0}
+ a^2(\tau) \varphi_{0} + 
4q\langle \chi^2 \rangle \varphi_{0} & = & 0  \label{app0}
\end{eqnarray}
where $\Omega_{p}^{2}= p^{2}-\ddot{a}/a+a^{2}$.

The main contribution to the last term on the left-hand side of
(\ref{appphi}) comes from the low-momentum,
classical modes of $\chi$. The fluctuations of $\varphi$ are 
driven by a classical force and therefore become classical themselves. 
This quantum-to-classical transition happens in the linear regime 
with respect to $\delta\varphi$. In this respect, it is similar to the one 
for the low-momentum modes of $\chi$, even though it is due to 
the force-driven rather than parametric amplification. 
We obtain the initial distribution function 
for $\varphi_{\p}$ and $\dot{\varphi}_{\p}$ analogous to (\ref{ini}). 
In fact, within the accuracy
of the classical approximation, we could have taken
the initial values for $\varphi$ and $\dot{\varphi}$ to be zeroes:
the classical force would produce fluctuations of $\varphi$ anyway.

The amplified fluctuations of $\varphi$, together with the 
low-momentum fluctuations of $\chi$ form a classical force for
the higher-momentum fluctuations of $\chi$, see (\ref{appchi}).
In this way, the power spectra propagate to larger momenta.

According to (\ref{appphi}), the $\delta\varphi$
subsystem is a resonator of high quality (the universe already
expands slowly at this stage). Therefore, the rapid growth
of $\delta\varphi$ will continue until
$\delta\varphi \sim \bar{\varphi}$,
and the approximation (\ref{appchi})--(\ref{appphi}), based on
(\ref{ll}), breaks down.\scite{resc} A more precise condition for 
the end of the rapid growth of $\delta\varphi$, obtained from 
the simulations, is
\begin{equation}
\langle(\delta \varphi)^2 \rangle \sim 0.1 \bar{\varphi}^2
\label{until}
\end{equation}
The system then enters the fully non-linear stage.

\section{The fully non-linear calculation}
To include all non-linear effects, we simulate the complete
non-linear equations of motion (\ref{eqphi}), (\ref{eqchi})
as a {\em classical} non-linear problem with random initial
conditions for $\chi$ and $\delta\varphi$. The random initial
conditions are distributed according to
(\ref{disini}) and a similar expression for $\delta\varphi$.
The initial conditions for $\varphi_{0}$ are: $\varphi_{0}(0)=1$,
$\dot{\varphi}_{0}(0)=0$.
The simulations were done on $128^{3}$ lattices, with
$m^2=10^{-12} M^2_{\rm Pl}$, both for massless and for massive
$X$.\scite{resc} For massless $X$, we determined the evolution
of the scale factor $a(\tau)$ self-consistently, taking into
account the effect of the produced fluctuations in the Einstein
equations.
For massive $X$, we used the matter-dominated form of $a(\tau)$.
Simulations were done also for the conformally-invariant model,
discussed below.

\begin{figure}
\psfig{file=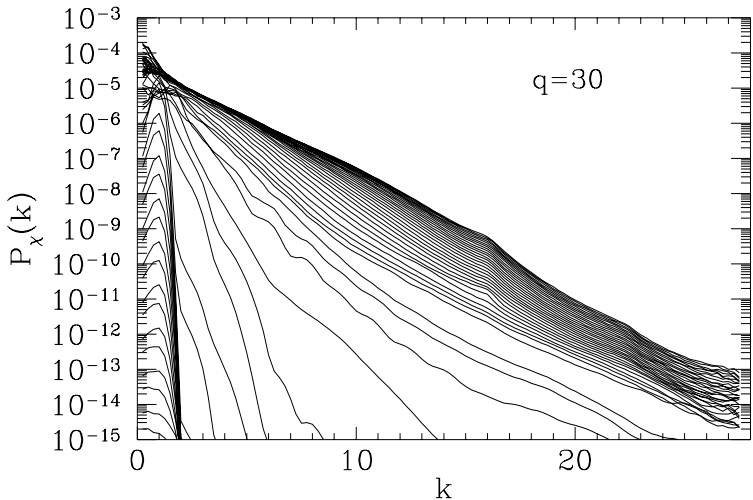,height=7cm,width=10.5cm}
\caption{Power spectrum of the field $\chi$
in the conformally invariant model, output every half-period
at the extrema of $\varphi_0(\tau)$.}
\label{Pws_chi_q30_lin_log}
\end{figure}

Additional results of the simulations are given in
Figs. \ref{xi2_phi2_qe4_exp}, \ref{Pws_chi_q30_lin_log},
\ref{xi2_phi2_qe6_m0}.
The variances and the power spectra are
defined according to (\ref{varlat}), (\ref{latphi}), (\ref{pws}).

We used the classical distribution (\ref{disini}) for all momenta
present on the lattice. This is possible because, even at the largest 
times that we consider, the main contributors to the physical 
quantities are modes with large occupation numbers. 
In general, a lattice calculation of the power spectrum can be trusted 
not up to the maximal momentum $k_{\rm max}$ present on the
lattice, but only up to the Nyquist momentum
$k_{\rm Ny}$; on a cubic lattice $k_{\rm Ny}=k_{\max}/\sqrt{3}$.
For example, in Fig. \ref{Pws_chi_q30_lin_log} (obtained for the 
conformally-invariant model), $k_{\rm max}\approx 28$ and 
$k_{\rm Ny}\approx 16$. For $3<k<16$, the latest outputs of the power 
spectrum are close to a perfect straight line in the log-linear plot:
\begin{equation}
P_{\chi}(k)\approx P_{0}\exp(-k/k_{0})
\label{exp}
\end{equation}
Such exponential fronts are obtained also for the power 
spectrum of $\varphi$ and the fields in other 
models.\scite{us,resc,grav}
Integrals of modest powers of $k$  computed with the power 
spectrum (\ref{exp}) will be saturated at $k\sim k_{0}$, where
$P_{\chi}(k)\sim P_{0}$, and the occupation numbers are large.

This distinguishes the highly non-equilibrium states emerging
after the inflaton's resonant decay from the equilibrium, thermal 
states. For a real  scalar field with a time-independent dispersion 
law $\epsilon(k)$, the occupation numbers are given by the Bose
distribution $n_{B}(k)$, and the power spectrum, at temperature $T$, 
is
\begin{equation}
P_{T}(k)=(2\pi)^{-3}\epsilon^{-1}(k) n_{B}(k)
=\frac{(2\pi)^{-3}\epsilon^{-1}(k)}{1-\exp[\epsilon(k)/T]}
\label{bose}
\end{equation}
The variance subtracted at $T=0$ is
\begin{equation}
\langle \chi^{2}\rangle_{T}-\langle \chi^{2} \rangle_{0}=
\int d^{3} k P_{T}(k)
\label{norm}
\end{equation}
The modes with $\epsilon(k)\ll T$ are essentially classical but 
the main contributors to (\ref{norm}) are the quantal modes, 
those with $n_{B}\sim 1$.

The new states are to a good accuracy classical, precisely
because they are highly non-thermal, with the power
spectra behaving like (\ref{exp}). For these states,
there is no ultraviolet catastrophe (no Rayleigh-Jeans problem).
Eventually, the system will thermalize, and will be 
dominated by quantum modes. Nevertheless, there is a prolonged 
stage in the evolution when the power spectra are concentrated 
at small momenta, and the system is essentially classical.

The non-linear phenomena described by the classical equations
of motion are creation, scattering, decay, and annihilation of
the classical waves arising through the amplification of 
quantum fluctuations. We refer to these phenomena as
rescattering.\scite{us}
In the language of particle physics, rescattering can be viewed
as stimulated creation, scattering, decay and annihilation 
of ``particles" in states with large occupation numbers.
A good analogy would be the scattering of 
waves from throwing pebbles in a pond, if those waves could be
made high enough for their non-linear interactions (rather than merely
interference) to become important. Such non-linear interactions
of waves occur in plasma physics and fluid dynamics.\scite{ZMR}

\section{Approximate methods for the non-linear stage}
\subsection{The Hartree approximation}
We will see below that as long as
\begin{equation}
\langle(\delta\varphi)^{2}\rangle \ll 
\bar{\varphi}^2/q_{\rm eff}
\label{1/qeff}
\end{equation}
we can neglect the fluctuations of $\varphi$ altogether.
For large $q_{\rm eff}$,  (\ref{1/qeff}) is a stronger condition 
than (\ref{ll}). In terms of the physical field $\phi$, (\ref{1/qeff})
reads
\begin{equation}
\langle(\delta\phi)^2\rangle \ll \phi_0^2(0)/q 
\label{1/q}
\end{equation}
We then get a system of two equations
\begin{eqnarray}
{\ddot \chi}_{\k}+\omega^{2}_{k}(\tau)\chi_{\k}
& = & 0 \label{hchi} \\
{\ddot \varphi}_{0} - ({\ddot a}/a) \varphi_{0}
+ a^2(\tau) \varphi + 
4q\langle \chi^2 \rangle \varphi & = & 0 \label{hphi}
\end{eqnarray}
These comprise the {\em Hartree
approximation} for the present model.\scite{KLS,Betal}
The classical average in (\ref{hphi}) approximates the corresponding
quantum average\scite{Betal} with the accuracy $O(1/n_{\rm amp})$, 
where $n_{\rm amp}$ is the typical occupation number in 
the amplified modes.

The equation (\ref{hchi}) is formally identical to (\ref{equ}), so
the operator solution (\ref{ope}) still applies, although the form
of $\varphi_{0}$ and, consequently, of $f_{k}$ will now be different.
The phase $\theta_{k}$, see (\ref{phase}), becomes almost 
time-independent under the same conditions (\ref{cond}) and $n_{k}\gg 1$. 
So, as long as $q_{\rm eff}\gg 1$, the variance
of $\chi$ in the Hartree approximation has the high-frequency ``beard''. 

\begin{figure}
\psfig{file=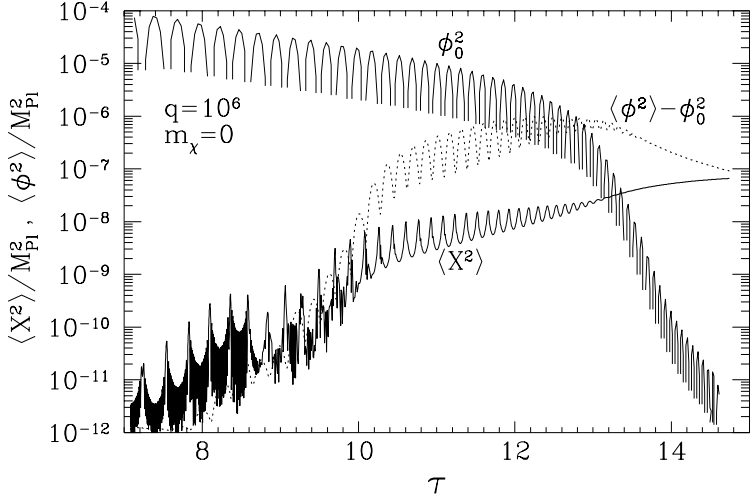,height=7cm,width=10.5cm}
\caption{Same as in Fig. 1 but for $q=10^6$, $M_X=0$.}
\label{xi2_phi2_qe6_m0}
\end{figure}

The time $\tau_{\rm sc}$ when the ``beard'' disappears on our pictures
while $q_{\rm eff}\gg 1$ (e.g. $\tau\approx 10.5$ in  
Fig. \ref{xi2_phi2_qe6_m0}) is the time when the Hartree approximation 
breaks down, and rescattering becomes a dominant effect.

\subsection{Corrections to the Hartree approximation}
To derive the condition (\ref{1/qeff}), we need
to estimate corrections to the Hartree approximation and see when 
they become large.
While (\ref{ll}) still applies, we can use for that purpose
the approximate equations (\ref{appchi})--(\ref{appphi}). 
We will do the estimate for the general case of a field $\chi$
with $N$ real components. The components will be labeled by
Greek indices $\alpha$, $\beta$,\ldots; a repeated index
is summed over and is sometimes omitted.

Eq. (\ref{appphi}) is solved by ($\p\neq 0$)
\begin{equation}
\varphi_{\p}(\tau) = \varphi_{\p}^{(0)} (\tau)+
4q\int_{0}^{\tau} d\tau' G_{p}(\tau,\tau') \varphi_{0}(\tau') 
\int d^{3}k \chi_{\k\alpha}^{*} \chi_{\k+\p,\alpha} (\tau')
\label{solphi}
\end{equation}
where $\varphi_{\p}^{(0)}$ is the solution to the free 
equation (\ref{ldphi}), and  $G_{p}(\tau,\tau')$ is the retarded
Green function for (\ref{ldphi}):
$G_{p}(\tau,\tau')\approx-\Omega_{p}^{-1}(\tau)
\sin[\omega_{p}(\tau-\tau')]$,
where $\Omega_{p}^{2}\approx p^{2}+a^{2}$.
The magnitude of $\varphi_{\p}^{(0)}$ is of the order of the initial 
condition for $\varphi_{\p}$ and is very small. At the stage when
the fluctuations of $\chi$ still rapidly grow, the main contribution
to the time integral in (\ref{solphi}) comes from an interval of 
duration of order $a^{-1}$ near $\tau'=\tau$, when the $\chi$ fluctuations 
are the largest. We then estimate (\ref{solphi}) as
\begin{equation}
\varphi_{\p}(\tau) \sim \frac{q {\bar \varphi}(\tau) }{a^{2}(\tau)}
\int d^{3}k \chi_{\k\alpha}^{*} \chi_{\k+\p,\alpha} (\tau)
\label{estphi}
\end{equation}
where we have used the estimate\scite{resc} $p\sim a(\tau)$ for
the typical momentum of the $\varphi$ fluctuations, so that also
$\Omega_{p}\sim a(\tau)$.

Eq. (\ref{appchi}) is solved by 
\begin{equation}
\chi_{\k\beta}(\tau) = \chi_{\k\beta}^{(0)}(\tau) + 
8q\int_{0}^{\tau} d\tau' F_{k}(\tau,\tau') 
\varphi_{0}(\tau') \int d^{3}p \varphi_{\p}^{*} \chi_{\k+\p,\beta}
\label{solchi}
\end{equation}
where $\chi_{\k\beta}^{(0)}(\tau)$ solves Eq. (\ref{hchi}) of the Hartree 
problem (\ref{hchi})--(\ref{hphi}),
and $F_{k}(\tau,\tau')$ is the retarded Green function for (\ref{hchi}); 
$F_{k}(\tau=\tau')=0$, ${\dot F}_{k}(\tau=\tau')=-1$. 
The integral over time in (\ref{solchi}) is estimated in the same 
manner as the one in (\ref{solphi}) 
\begin{equation}
\frac{{\bar \varphi}(\tau)}{a \omega_{k}(\tau)} \int d^{3}p 
\varphi_{\p}^{*} \chi_{\k+\p,\beta}
\label{estchi}
\end{equation}
where $\omega_{k}(\tau)\sim \sqrt{q}{\bar \varphi}(\tau)$.

Using the estimates (\ref{estphi}) and (\ref{estchi}) in 
(\ref{solchi}) we obtain
\begin{equation}
\chi_{\k\beta}-\chi_{\k\beta}^{(0)} \sim
\frac{q^{2}{\bar \varphi}^{2}}{a^{3}\omega_{k}}
\int d^{3} k' d^{3} p \chi^{*}_{\k'+\p,\alpha} \chi_{\k',\alpha} 
\chi_{\k+\p,\beta}
\label{cor}
\end{equation}
We recall at this point that $\chi$ and $\varphi$ are random 
variables. By the translational invariance and the $O(N)$ symmetry,
\begin{equation}
\langle \chi_{\k\alpha}^{*} \chi_{\k'\beta} \rangle = 
N^{-1}\langle \chi_{\k}^{*} \chi_{\k} \rangle 
\delta_{\alpha\beta} \delta_{\k\k'}
\rightarrow (2\pi)^{3} N^{-1} V^{-1}
\langle \chi_{\k}^{*} \chi_{\k} \rangle
\delta_{\alpha\beta} \delta(\k-\k')
\label{tran}
\end{equation}
So, for  instance, the average of $\varphi_{\p}$ ($\p\neq 0$)
over realizations of the initial data is zero. 

The first correction  to the Hartree approximation is obtained
when we replace $\chi$ on the right-hand sides of 
(\ref{solphi}), (\ref{solchi}) with $\chi^{(0)}$. 
In the Hartree approximation, different pairs of modes of $\chi$ with
the opposite momenta ($\k$ and $-\k$) are statistically independent.
For example, for any $\p\neq 0$
\begin{eqnarray}
\lefteqn{\langle
\chi^{*(0)}_{\k\beta} \chi^{*(0)}_{\k'+\p,\alpha} 
\chi^{(0)}_{\k'\alpha'}  \chi^{(0)}_{\k+\p,\beta'}
\rangle  = } \nonumber \\ 
& & 
\langle \chi^{*(0)}_{\k\beta} \chi^{*(0)}_{\k'+\p,\alpha} 
\rangle \langle 
\chi^{(0)}_{\k'\alpha'}  \chi^{(0)}_{\k+\p,\beta'} \rangle +
\langle \chi^{*(0)}_{\k\beta} \chi^{(0)}_{\k'\alpha'} \rangle
\langle \chi^{*(0)}_{\k'+\p,\alpha} \chi^{(0)}_{\k+\p,\beta} 
\rangle 
\label{gau}
\end{eqnarray}
for any $\p\neq 0$. Higher-order correlators decompose in a 
similar way.

Using this approximation in (\ref{estphi}), we estimate the 
mean square of $|\varphi_{\p}|$  as
\begin{equation}
\langle |\varphi_{\p}|^{2}\rangle \sim 
\frac{q^{2}   {\bar \varphi}^{2}}{a^{4}NV}\int d^{3} k 
\langle |\chi_{\k}|^{2} \rangle 
\langle |\chi_{\k+\p}|^{2} \rangle  
\label{msphi}
\end{equation}
and the variance $\langle(\delta\varphi)^{2}\rangle=
V^{-1}\int d^{3}p |\varphi_{\p}|^{2}$ as
\begin{equation}
\langle(\delta\varphi)^{2}\rangle \sim q^{2}   {\bar \varphi}^{2} 
\langle \chi^{2} \rangle_{\rm v}^{2}/a^{4}N
\label{varphi}
\end{equation}
We can similarly estimate the first correction to 
$\langle \chi^{*}_{\k}\chi_{\k}\rangle$. Towards the end 
of the Hartree stage, it is of order of the average square of 
the right-hand side of (\ref{cor}). 
(The cross correlation between the right-hand side and 
$\chi^{(0)}$ is at these times a smaller effect for large $N$.) 
We find that the first correction becomes of the order of 
the leading term when\scite{resc}
\begin{equation}
\langle \chi^{2} \rangle_{\rm v} 
\sim \sqrt{N}~\frac{a^{3}(\tau)}{q^{3/2}{\bar \varphi}}
{}~~~~~~~\mbox{i.e.}
{}~~~\langle X^{2} \rangle_{\rm v} 
\sim \sqrt{N}~\frac{\phi_{0}^{2}(0)}{q q_{\rm eff}^{1/2}}
\label{varchi}
\end{equation}
According to (\ref{varphi}), at that time
\begin{equation}
\langle(\delta\varphi)^{2}\rangle \sim \frac{a^{2}}{q}
{}~~~~~~~\mbox{i.e.}
{}~~~\langle \phi^{2} \rangle \sim \phi_{0}^{2}(0)/q
\label{sim1/q}
\end{equation}
cf. the condition (\ref{1/q}).

The estimate (\ref{varchi}) grows indefinitely with $N$, for
a fixed $q$. The physical variance cannot do that.
So, for sufficiently large $N$, $N>N_{0}(q)$, where $N_{0}$
increases with $q$, the Hartree approximation does not break
down at all during the rapid growth of $\langle \chi^{2}\rangle$,
i.e. the condition (\ref{sim1/q}) is never reached.

Otherwise,
the estimate (\ref{sim1/q}) determines the time $\tau_{\rm sc}$
after which rescattering
begins to strongly influence the evolution of $\langle\chi^{2}\rangle$.
Because in deriving (\ref{sim1/q}) we neglected various numerical factors, 
it should be more of a scaling law with respect to $q$ than of an actual 
numerical estimate. Nevertheless, it turns out to correspond
numerically quite well to $\langle(\delta\varphi)^{2}\rangle$ at 
the time when the ``beard'' disappears on our pictures, whenever
that happens while $q_{\rm eff}\gg 1$.
For example, in Fig. \ref{xi2_phi2_qe6_m0}, the beard disappears at
$\tau=\tau_{\rm sc}\approx 10.5$. At that time, 
$\langle(\delta\phi)^{2}\rangle\sim 10^{-7} M_{\rm Pl}^2$, in agreement
with (\ref{sim1/q}).  
On the other hand, in Fig. \ref{xi2_phi2_qe4_exp}, $q_{\rm eff}$
becomes of order 1 already at $\tau\sim 10$, and even the maximal value
of $\langle(\delta\phi)^{2}\rangle$ is somewhat smaller than the estimate
(\ref{sim1/q}). 

At $\tau>\tau_{\rm sc}$, the system exhibits chaotic 
behavior, characteristic of a non-linear classical system. The stage when
\begin{equation}
\phi_{0}^{2}(0)/q \lsim \langle(\delta\phi)^{2}\rangle \ll 
0.1 {\bar \phi}^{2}
\label{chao}
\end{equation}
has been called the {\em chaotization} stage.\scite{resc,grav}
When the second condition in (\ref{chao}) is broken, the system enters
the fully non-linear stage with developed chaotic behavior.
After chaotization, the system is in a slowly evolving (quasi-steady)
state with smooth power spectra,
see e.g. Fig. \ref{Pws_chi_q30_lin_log}. Such states are naturally called
{\em turbulent}. It is the chaotic evolution that finally establishes 
the maximal values attained by the variances of the fields.

The turbulent states emerge in a variety of models of the inflaton's
resonant decay (cf. the conformally invariant case below).
The power spectra in these states in general consist of a segment of 
a power law
at the smallest $k$, $k\sim 1$, from where the exponential front
(\ref{exp}) takes off. We expect that at later times, the front will
slowly move to larger momenta, and the region of $k$ occupied by 
the power law will grow.

\section{Decay of the inflaton}
An interesting feature of the massive model in the 
expanding universe is that the inflaton decays completely
fairly soon after it starts to oscillate, see Figs. \ref{xi2_phi2_qe4_exp},
\ref{xi2_phi2_qe6_m0}. This early decay does 
not occur in the same model without the expansion or in the 
conformally invariant case (discussed below).

Because the  inflaton decays fast, and the expansion of the
universe is already slow, the total energy is almost conserved during
the decay. One can obtain a simple, albeit non-rigorous, estimate
of the maximal size of $X$ fluctuations from the energy conservation.
We have (in terms of the physical fields)
\begin{equation}
m^{2}{\bar\phi}^{2}(\tau_{1}) \sim 
g^{2} \langle(\delta\phi)^{2}\rangle(\tau_{2})
\langle X^{2} \rangle_{\max}
\label{encon}
\end{equation}
where $\tau_{1}$ is a moment right before the decay, and 
$\tau_{2}$ is the moment when $\langle X^{2} \rangle$ is maximal;
we have assumed that a significant portion of the inflaton's energy went 
into the $X$ fluctuations. Using 
$\langle(\delta\phi)^{2}(\tau_{1}) \rangle \sim 
0.1 \bar{\phi}^{2}(\tau_{1})$, 
see (\ref{until}), and assuming that $\langle(\delta\phi)^{2} \rangle$ 
does not change drastically during the decay, we get
\begin{equation}
\langle(\delta\phi)^{2} \rangle (\tau_{2}) \sim 
0.1 {\bar \phi}^{2}(\tau_{1})
\label{until1}
\end{equation}
From (\ref{encon}) and (\ref{until1}), $\langle X^{2} \rangle_{\max}
\sim m^{2}/g^{2}\sim \phi_{0}^{2}(0)/q$ (neglecting numerical factors), 
or
\begin{equation}
\langle X^{2} \rangle_{\max} = \epsilon \phi_{0}^{2}(0)/q
\label{max}
\end{equation}
where $\epsilon$ takes into account the fraction of the inflaton's
energy that went to $X$ and other numerical factors. 

The $1/q$ estimate (\ref{max}) agrees with the scaling of our data 
points and with an estimate obtained from the 
self-similarity hypothesis.\scite{resc} From the data, we
find $\epsilon\sim\mbox{0.1--1}$.

\section{Conformally-invariant model}
This model has the scalar potential
\begin{equation}
V(\phi,X)={1\over 4} \lambda\phi^{4} + {1\over 2} g^2 \phi^2 X^2
\label{pot1}
\end{equation}
The fields $\phi$ and $X$ are massless. For the inflaton self-coupling
$\lambda$ we take a realistic value $\lambda=10^{-13}$.

The inflaton oscillations start at $\phi_0(0)\approx 0.35 M_{\rm Pl}$ , 
$d{\phi}_{0}(0)/d\eta=0$;\scite{wide}
$\eta$ is the conformal time: $d\eta=dt/a(t)$.
The rescaled conformal variables are
\begin{center}
\begin{tabular}{ll}
$\tau  =  \sqrt{\lambda}\phi_{0}(0) \eta\; ;$ 
&  ~~~~~~~$\varphi= \phi a/\phi_{0}(0)$ \\
$\mbox{\boldmath $\xi$} =\sqrt{\lambda}\phi_{0}(0) \mbox{\bf x}\; ;$  
& ~~~~~~~$\chi=X a/\phi_{0}(0)$
\end{tabular} 
\end{center}
so that $\varphi_{0}(0)=1$. The universe is radiation dominated,
$a(\tau)/a(0)\approx 0.51\tau+1$.

The equations of motion in the conformal variables
\begin{eqnarray}
&&{\ddot \varphi} - \nabla_{\xi}^2 \varphi  + 
\varphi^3 +4q\chi^2\varphi= 0 \; ,
\nonumber \\ 
&&{\ddot \chi} - \nabla_{\xi}^2 \chi   +4q\varphi^2\chi= 0 \; .
\label{eqmc}
\end{eqnarray}
are the same as in the flat space-time. The resonance parameter
in this case is $q \equiv g^2/4\lambda$.

The power spectrum of $\chi$ for $q=30$ is shown in
Fig. \ref{Pws_chi_q30_lin_log}.
Notice the formation of the exponential tail, cf. (\ref{exp}). 
The latest outputs in Fig. \ref{Pws_chi_q30_lin_log}  fall almost 
exactly onto each other---a sign of a quasi-steady state.

\section{Conclusions}
Resonant decay of the inflaton leads to formation of highly
non-equilibrium states after inflation.
These states are dominated by interacting classical waves.
As we have shown elsewhere,\scite{grav}  collisions of these classical 
waves give rise to a potentially observable
background of relic gravitational waves. 

\section*{Acknowledgments}
I would like to thank the organizers of SEWM'97 for holding 
a stimulating conference. The author's work was supported
in part by the U.S. Department of Energy under Grant DE-FG02-91ER40681 
(Task B), by the National Science Foundation under Grant PHY 95-01458, 
and by the Alfred P. Sloan Foundation.

\end{document}